\documentclass[aps,amsmath,amssymb,prb,twocolumn,superscriptaddress,unsortedaddress]{revtex4}
\usepackage{epsf}
\usepackage{graphicx}
\usepackage{amsfonts}
\usepackage{sidecap}
\sidecaptionvpos{figure}{c}

\begin{document}

\title{Topological surface electronic states in candidate nodal-line semimetal CaAgAs}

\author{Xiao-Bo~Wang}
\affiliation {Department of Physics, Southern University of Science and Technology (SUSTech), Shenzhen, Guangdong 518055, China}

\author{Xiao-Ming~Ma}
\affiliation {Department of Physics, Southern University of Science and Technology (SUSTech), Shenzhen, Guangdong 518055, China}
\affiliation {Beijing National Laboratory for Condensed Matter Physics, and Institute of Physics, Chinese Academy of Sciences, Beijing 100190, China}

\author{Eve~Emmanouilidou}
\affiliation {Department of Physics and Astronomy and California NanoSystems Institute, University of California, Los Angeles, California 90095, USA}

\author{Bing~Shen}
\affiliation {Department of Physics and Astronomy and California NanoSystems Institute, University of California, Los Angeles, California 90095, USA}

\author{Chia-Hsiu~Hsu}
\affiliation {Department of Physics, Southern University of Science and Technology (SUSTech), Shenzhen, Guangdong 518055, China}

\author{Chun-Sheng~Zhou}
\affiliation {Department of Physics, Southern University of Science and Technology (SUSTech), Shenzhen, Guangdong 518055, China}

\author{Yi~Zuo}
\affiliation {Department of Physics, Southern University of Science and Technology (SUSTech), Shenzhen, Guangdong 518055, China}

\author{Rong-Rong~Song}
\affiliation {Department of Physics, Southern University of Science and Technology (SUSTech), Shenzhen, Guangdong 518055, China}

\author{Su-Yang~Xu}
\affiliation {Department of Physics, Massachusetts Institute of Technology, Cambridge, Massachusetts 02139, USA}

\author{Gan~Wang}
\affiliation {Department of Physics, Southern University of Science and Technology (SUSTech), Shenzhen, Guangdong 518055, China}

\author{Li~Huang}
\affiliation {Department of Physics, Southern University of Science and Technology (SUSTech), Shenzhen, Guangdong 518055, China}

\author{Ni~Ni}
\email{nini@physics.ucla.edu}
\affiliation {Department of Physics and Astronomy and California NanoSystems Institute, University of California, Los Angeles, California 90095, USA}

\author{Chang~Liu}
\email{liuc@sustc.edu.cn}
\affiliation {Department of Physics, Southern University of Science and Technology (SUSTech), Shenzhen, Guangdong 518055, China}

\date{\today}

\clearpage

\begin{abstract}

We investigate systematically the bulk and surface electronic structure of the candidate nodal-line semimetal CaAgAs by angle resolved photoemission spectroscopy and density functional calculations. We observed a metallic, linear, non-$k_z$-dispersive surface band that coincides with the high-binding-energy part of the theoretical topological surface state, proving the topological nontriviality of the system. An overall downshift of the experimental Fermi level points to a rigid-band-like $p$-doping of the samples, due possibly to Ag vacancies in the as-grown crystals.

\end{abstract}


\maketitle

The discovery of time reversal invariant topological insulators has dominated the field of condensed matter physics over the past decade. Unique nontrivial topological properties in these fermionic systems relate concepts from high energy physics to various quasiparticle excitations in condensed matter, causing the systems to resist small perturbations due to the protection by particular topological invariants.\cite{Hasan_Review, Zhang_Review} In the so-called topological semimetals, such nontrivialities appear as the touching of valence and conduction bands at isolated points, closed lines or planes in the momentum space. Such materials exhibit chiral anomaly,\cite{Ong_Science, JiaShuang_ABJ} topological surface Fermi arcs,\cite{Suyang_TaAs, Ding_TaAs, HgCr2Se4, MultiWeyl, Suyang_FermiArc, TaAs_DFT, Shi_TaP, Weyl_Heusler, Weyl_Multilayer} and/or superconducting zero modes,\cite{Wang_Majorana, Leon_Majorana, Jia_Majorana} whose quasiparticle excitations correspond directly to the Dirac, Weyl, and Majorana fermions. These quasiparticles differ from the actual entities in high energy physics, but obey the same underlying principles in quantum field theory, offering the opportunity to investigate the fundamental physical laws that govern a large subset of quantum condensed matter as well as creating a new approach for developing a broad range of low-power, high efficiency spintronic and quantum computing devices.\cite{Hasan_Review, Zhang_Review, DasSarma_Review}

Topological nodal-line semimetals (NLSM) exhibit novel topological properties that are manifested by surface states in the form of a drum-like membrane living in the continuous toroidal isosurface gap in three dimensional (3D) momentum space. With strong spin-orbit coupling (SOC), the nodal lines in these materials are either protected by reflection or mirror symmetry, or gapped out due to the lack of such symmetries.\cite{Balents_NLSM, Fu_NLSM, Fang_NLSM, Rappe, Kawazoe} Recent examples/promising candidates of NLSM include Pb(Tl)TaSe$_2$ (Refs. \onlinecite{PbTaSe2, TlTaSe2}), ZrSiX (X = S, Se, Te) (\mbox{Refs.} \onlinecite{ZrSiS_NC, ZrSiS_PRB}), Ca$_3$P$_2$ (Ref. \onlinecite{Ca3P2}), CaTe (Ref. \onlinecite{CaTe}), Cu$_3$PdN (\mbox{Ref.} \onlinecite{Cu3PdN}), GdSbTe (Ref. \onlinecite{GdSbTe}), body-centered orthorhombic C$_{16}$ (Ref. \onlinecite{C16}), compressed black phosphorus (Ref. \onlinecite{B_Phosphorus}), etc. However, the band structures of these materials are often so complex that multiple irrelevant trivial or non-trivial \mbox{pockets} coexist with the drum-like surface states at the Fermi level, masking the quantum transport signals from the nodal lines. Experimentally realizing a clean, ``hydrogen-atom-like'' NLSM is therefore of urgent need. Single crystalline CaAgAs is theoretically predicted to be among the best candidate NLSMs since its theoretical Fermi surface contains no more than a circular nodal contour linked by the topological surface state,\cite{CaAgX_Japan} which gives rise to the ultra-low magneto-resistance found by transport measurements.\cite{NiNi} In this paper, we systematically investigate the NLSM state of CaAgAs which is solely protected by mirror reflection symmetry through a comparison between angle resolved photoemission spectroscopy (ARPES) measurements and density functional theory (DFT) calculations. Out DFT calculations show that without SOC, a topological nodal ring enclosing $\Gamma$ appears in the first Brillouin zone (BZ) near the theoretical Fermi energy. The strong SOC introduced mainly by heavy As atoms causes the opening of a substantial band gap at the line node, turning the system into a narrow-gap topological insulator. Our ARPES experiments show that metallic surface electronic states exist in this system, agreeing reasonably well with results from DFT calculations. A $\sim$500 meV downshift of the experimental Fermi level is indicative of effective $p$-doping due to Ag vacancies in the crystals. Such surface states provide a wonderful playground for exploration of exotic physical properties such as surface magnetism or superconductivity as well as long-range Coulomb interactions.\cite{Ca3P2, Kim_LL, Kim_Coulomb}

\begin{SCfigure*}
\centering
\includegraphics[width=13cm]{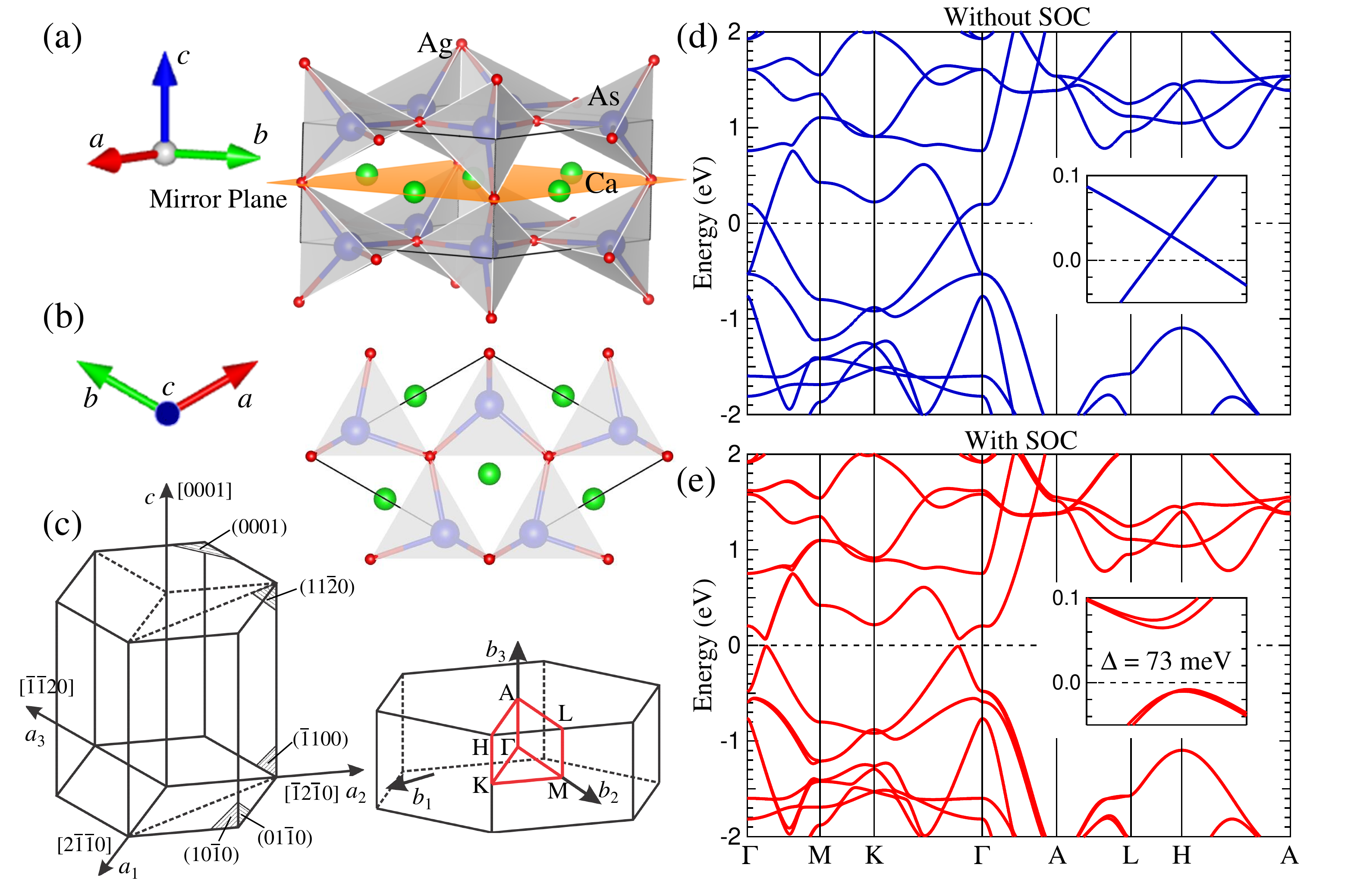}
\caption{(Color online) Crystal and first-principles band structure. (a) Crystal structure of CaAgAs. The AgAs$_4$ octahedra are shown in grey. The orange plane represents the mirror plane that protects the TSS. (b) Same as (a) but seen along the $c$ axis. (c) The Brillouin zone and notations of high symmetry points and planes. (d)-(e) DFT band structure of CaAgAs (d) without and (e) with spin orbit coupling (SOC). Insets show the detailed band dispersion along $\Gamma$-M near the nodal point. The nodal-line around $\Gamma$ is gapped out under SOC, yielding a topological insulator with $\Delta = 73$ meV.}\label{Background}
\end{SCfigure*}

Single crystals of CaAgAs were grown by the flux method described in Ref. \onlinecite{NiNi}. ARPES measurements were performed at the ``Dreamline'' beamline of the Shanghai Synchrotron Radiation Facility (SSRF) with a Scienta DA30 analyzer. The samples were cleaved $in~situ$ and measured at 15 K in a vacuum better than $7 \times 10^{-11}$ mbar. The energy resolution was set at 15 meV for the Fermi surface mapping, and the angular resolution was set at $0.1^\circ$. The ARPES data were collected using linearly horizontal-polarized lights with a vertical analyzer slit. Electronic structures at the top, (0001) and the side, $(11\bar{2}0)$ planes are achieved by cleaving the crystals along the corresponding pre-selected surfaces. Our first-principles DFT calculations were performed by the Vienna $Ab$-$initio$ Simulation Package (VASP) based on the generalized gradient approximation of the Perdew-Burke-Ernzerhof (PBE) type and the projector-wave (PAW) pseudopotentials. The energy cutoff was set to be 520 eV for the plane-wave basis, and the Brillouin-zone integration was performed on a Monkhorst-Pack $\Gamma$-centered $8 \times 8 \times 14$ $k$-point mesh.\cite{Ab_initio93, Plane-Wave, PAW} The angle-resolved density of states (AR-DOS) was processed by the Wannier90 package; the hopping matrix of the Hamiltonian is obtained by the maximally localized Wannier functions (MLWFs).\cite{Wannier} Using these optimized MLWFs, the surface Green's function of the semi-infinite lattice is constructed via the recursive method until the final convergence is reached.\cite{Rubio_Green}

Topological surface states (TSSs) of the NLSMs are generally protected by the combination of inversion, time reversal and the corresponding spin SU(2) symmetries. In some special cases like CaAgAs, these states are only protected by mirror reflection symmetry, and the material is characterized as a type-A NLSM.\cite{Fang_NLSM} CaAgAs crystalizes in a hexagonal pyrochlore-type structure that belongs to the space group $P\bar{6}2m$ (No. 189) [Fig. \ref{Background}(a),(b)]. The volume of the primitive cell is $V = 220.1$ $\mathrm{\AA}^3$ with lattice constants $a = b = 7.201$ $\mathrm{\AA}$, $c = 4.263$ $\mathrm{\AA}$. A mirror reflection plane located at the calcium plane intersecting \mbox{two} layers of AgAs$_4$ octahedras [orange plane in Fig. \ref{Background}(a)] protects the drum-head TSS.\cite{CaAgX_Japan, NiNi} Figs. \ref{Background}(d),(e) demonstrate the DFT bulk band structure of CaAgAs without and with SOC, respectively. Without SOC, CaAgAs is a NLSM with two gapless bands having opposite mirror parity on opposite sides of the $\Gamma$ point near the Fermi level [Fig. \ref{Background}(d)]. These two bands cross each other due to band inversion and form a nodal ring around the $\Gamma$ point. When the nodal ring is projected onto the (0001) surface, it produces topologically protected surface states on either the inside or the outside of the ring, depending on the termination of the cleavage surface.\cite{CaAgX_Japan} When SOC is introduced in the system, a 73-meV gap opens along the nodal line of the Dirac-like bulk band, and the system becomes a topological insulator [Fig. \ref{Background}(e)] whose in-gap TSS is now protected by time reversal symmetry.

\begin{figure}[t]
\centering
\includegraphics[width=8.9cm]{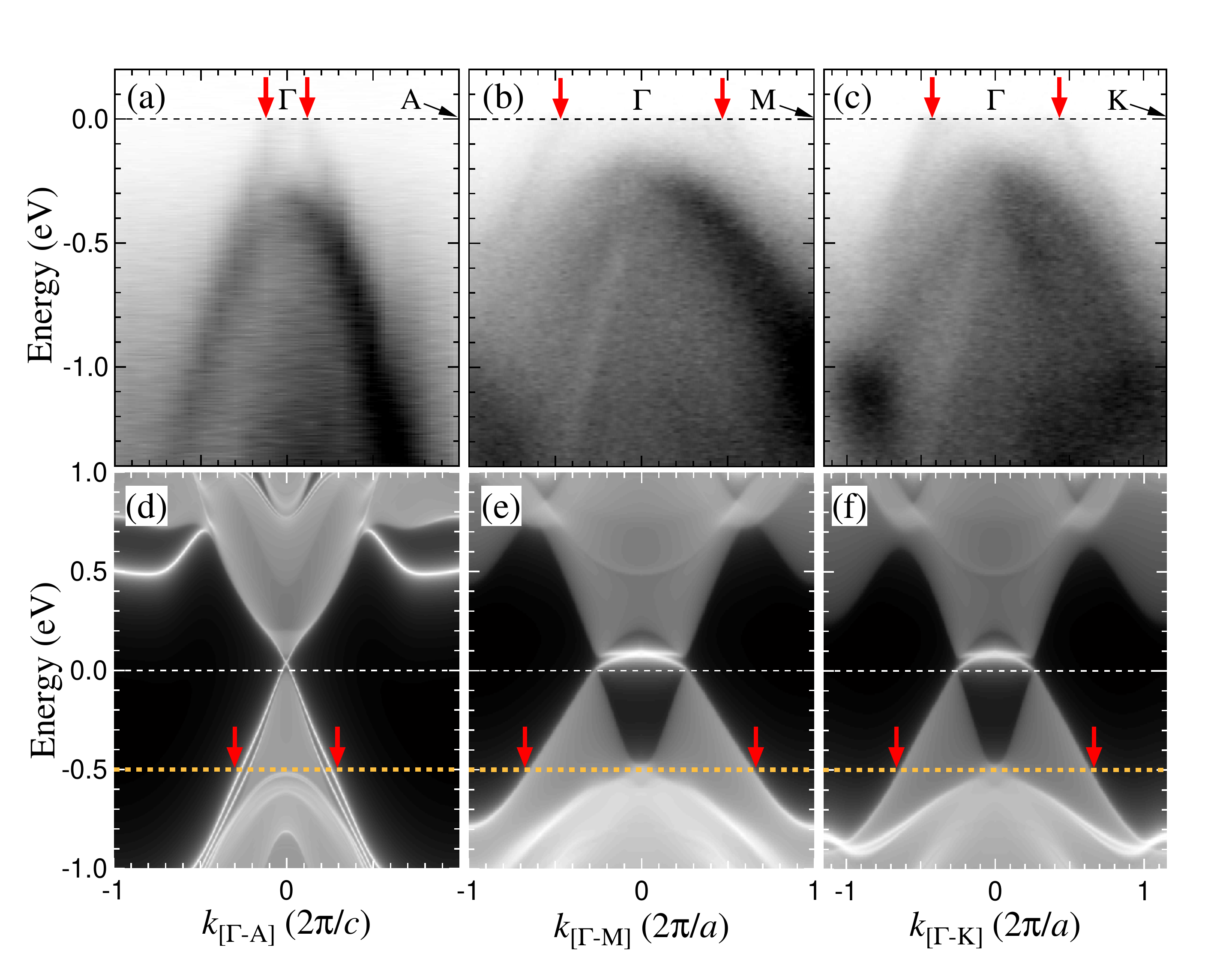}
\caption{(Color online) ARPES and DFT band dispersion along high symmetry directions. Red arrows denote the lower part of the TSS we predicted by DFT calculations and observed by ARPES. (a)-(c) ARPES $k$-$E$ maps along three high symmetry directions (a) $\Gamma$-A, (b) $\Gamma$-M and (c) $\Gamma$-K. (d)-(f) DFT-derived angle-resolved density of states along three high symmetry directions (d) $\Gamma$-A, (e) $\Gamma$-M and (f) $\Gamma$-K. Brighter curves denote the topological surface states. The yellow dashed line shows an approximate Fermi level position that agrees with the ARPES data.}\label{Bands}
\end{figure}

Our ARPES data proves the topological \mbox{nontriviality} of the system by observing unambiguously a TSS of \mbox{CaAgAs}. This is shown in Fig. \ref{Bands} where we investigate the ARPES $k$-$E$ dispersion of the bulk and surface \mbox{bands} along the high symmetry directions [Figs. \ref{Bands}(a)-(c)], in comparison with our calculated AR-DOS [Figs. \ref{Bands}(d)-(f)]. In Fig. \ref{Bands}(a), we observe linear, relatively sharp bands (red arrows) at both sides of the $\Gamma$ points along the $\Gamma$-A direction. These bands do not exist in the bulk DFT calculations [Fig. \ref{Background}(e)] but are present in the corresponding AR-DOS [Fig. \ref{Bands}(d)], which is obtained from the surface Green's functions containing information from both the bulk and the surface bands (since a semi-infinite lattice is used in the calculation). Along $\Gamma$-M and $\Gamma$-K [Figs. \ref{Bands}(b)-(c)], linear bands (red arrows) also appear at both sides of the $\Gamma$ points, along the edge of the $\Gamma$-centered intensity that comes mainly from the crystal bulk, which again coincide with the bright surface state we obtained by DFT calculations [red arrows in Figs. \ref{Bands}(e)-(f)] that is connected to the X-shaped Dirac cone TSS near the theoretical chemical potential. Therefore, these linear bands we observed by ARPES comprise the TSS of CaAgAs beyond doubt. Note that there are two possible surface terminations in this crystal: Ca$_3$As and Ag$_3$As$_2$. The AR-DOS is calculated for both terminations (not shown), but the ARPES dispersion is found to be analogous to the bands in the Ca$_3$As surface, meaning that the Ca$_3$As atomic plane is the natural cleavage surface of CaAgAs in the (0001) direction. This result points out that the observed linear \mbox{states} are the (partially observed) TSS that stem from the topological insulating Dirac cone. Without SOC, this state would change to the drum-head TSS in a NLSM. Therefore, the dominant feature of the predicted nodal ring near the theoretical Fermi level has been demonstrated by our ARPES data and first-principles calculation results.

\begin{figure}[t]
\centering
\includegraphics[width=8.8cm]{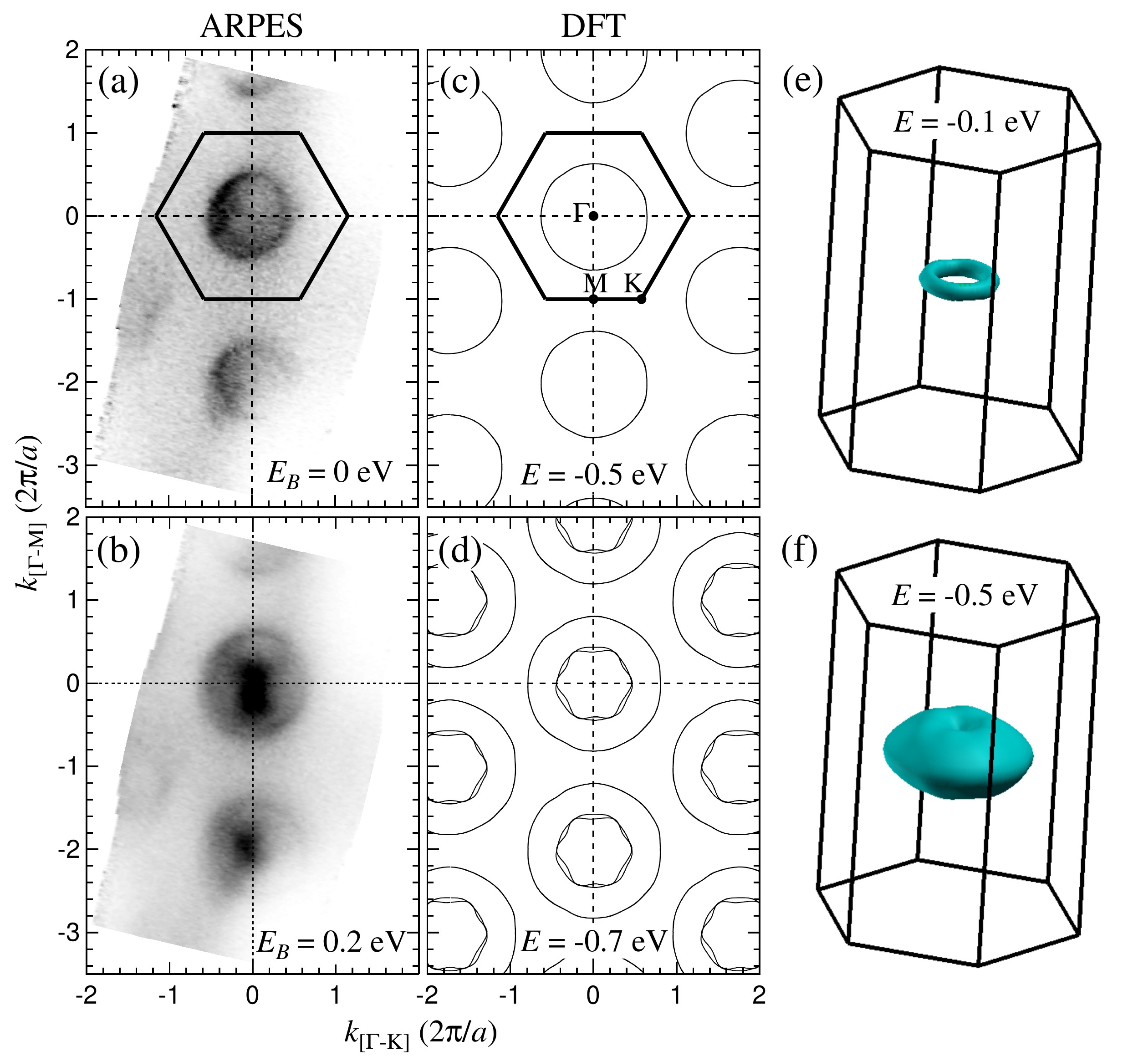}
\caption{(Color online) ARPES and DFT constant energy contours at the (0001) plane. (a)-(b) ARPES maps at (a) the Fermi level and (b) 0.2 eV binding energy using 70 eV photons. (c)-(d) DFT bulk bands at binding energies of (c) 0.5 eV and (d) 0.7 eV. (e)-(f) 3D constant energy contour obtained by DFT calculations at binding energies of (e) 0.1 eV and (f) 0.5 eV.}\label{Top_Fermi}
\end{figure}

\begin{SCfigure*}
\centering
\includegraphics[width=12cm]{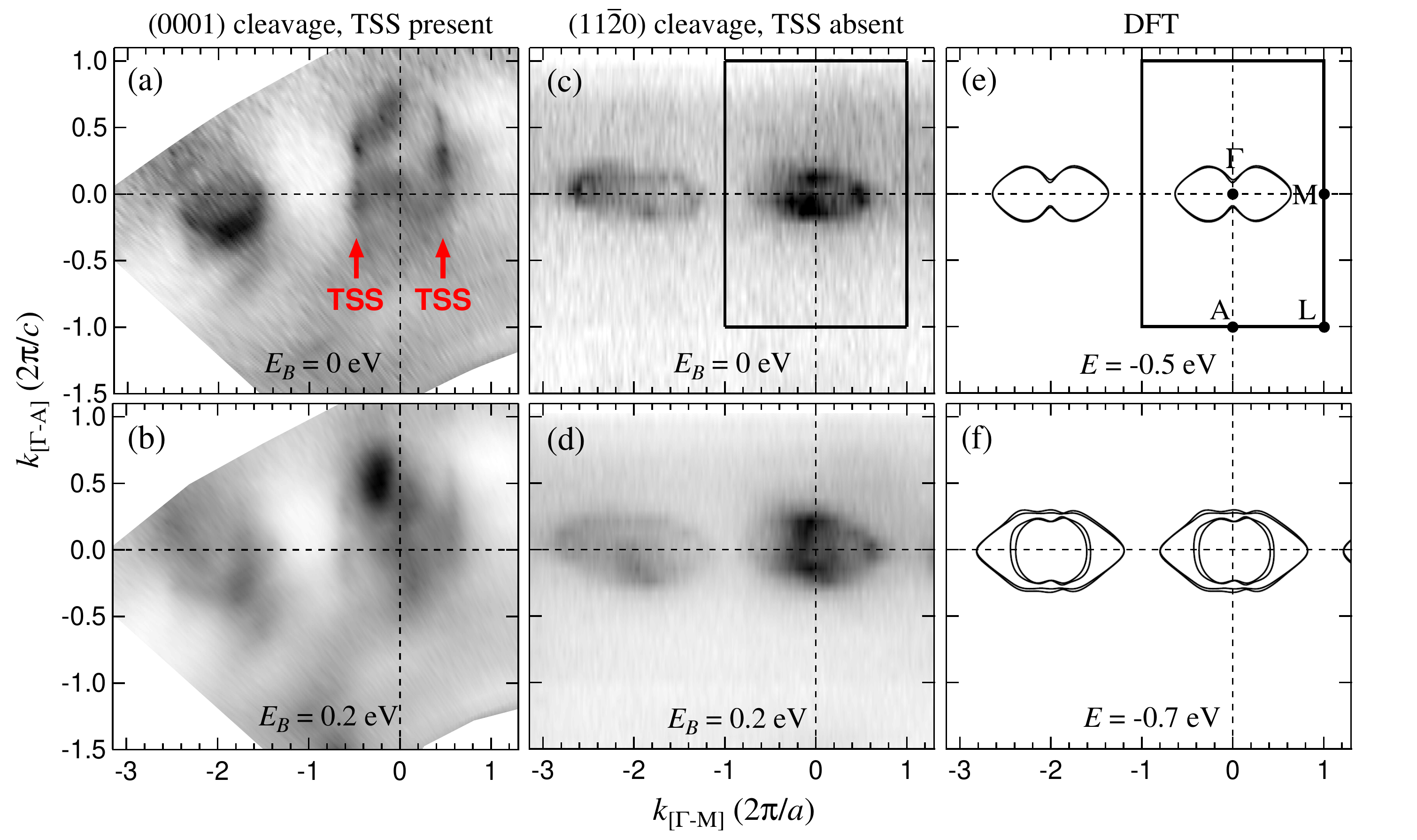}
\caption{(Color online) ARPES and DFT constant energy contours at the ($11\bar{2}0$) plane. (a)-(b) ARPES map for a sample cleaved along the (0001) plane at binding energies of (a) 0 eV and (b) 0.2 eV. Topological surface states (TSS, re
d arrows) appear with such cleaving condition. (c)-(d) Same as (a)-(b) but for a sample cleaved along the ($11\bar{2}0$) surface. No TSS can be seen with such cleaving condition. (e)-(f) Bulk constant energy cuts by DFT calculations at energies of (a) -0.5 eV, (b) -0.7 eV.}\label{kz_TSS}
\end{SCfigure*}

Fig. \ref{Top_Fermi} shows the ARPES constant energy maps obtained at the (0001) top surface of CaAgAs at the experimental binding energies of 0 and 0.2 eV [Figs. \ref{Top_Fermi}(a),(b)], compared with the corresponding DFT results at the calculated binding energies of 0.5 and 0.7 eV [Figs. \ref{Top_Fermi}(c),(d)]. The sample here is also cleaved along the (0001) surface, in which case the TSS, if present, will be projected onto the surface and be seen in both the constant energy maps and the $k$-$E$ cuts. We observe from Fig. \ref{Top_Fermi}(a)-(d) that the (0001) Fermi surface of CaAgAs contains only a single circular band enclosing $\Gamma$, which is in good \mbox{agreement} with the calculated bulk isosurface cut at the energy value of 0.5 eV below the calculated Fermi level. At experimental and theoretical binding energies of 0.2 and 0.7 eV respectively, the maps also show similar band features, containing \mbox{two} $\Gamma$-centered circular rings with additional structure on the inner ring. This comparison indicates that, though the ARPES and DFT bands have one-to-one correspondence, the experimental Fermi level is about 0.5 eV lower than the theoretical one, meaning that the real system is substantially $p$-doped. Fig. \ref{Top_Fermi}(e),(f) show the 3D DFT constant energy contours at binding energies of 0.1 and 0.5 eV, respectively. The 0.5 eV contours capture the 3D bands at the experimental $E_F$ quite well, whereas the 0.1 eV one predicts a toroidal band dispersion in the $k$-space right below the nodal line of the system.

Fig. \ref{kz_TSS} shows the ARPES constant energy maps obtained at the $(11\bar{2}0)$ $side$ surface [see Fig. \ref{Background}(c) for definition] at $E_B = 0$ and 0.2 eV, compared with the DFT results at $E_B = 0.5$ and 0.7 eV. Two types of experimental configurations are used to obtain the ARPES band dispersions at this plane. In Figs. \ref{kz_TSS}(a) and (b), we cleaved the sample along the (0001) top surface, and measured the dispersion at the $(11\bar{2}0)$ side surface by varying the incoming photon energies. In this case, the bands are projected onto the (0001) top surface where the TSS is visible. In Figs. \ref{kz_TSS}(c) and (d), we cleaved the sample \mbox{along} the $(11\bar{2}0)$ side surface, and measured the dispersion at the $(11\bar{2}0)$ side surface by varying the measurement angle with respect to the sample surface normal (routine ARPES mapping). In this case, the bands are projected onto the $(11\bar{2}0)$ side surface where the TSS is absent. From the ARPES maps we indeed see that the two additional vertical lines beside the peanut-shaped bulk bands appear only in Figs. \ref{kz_TSS}(a),(b) but not in Figs. \ref{kz_TSS}(c),(d). These states cross the Fermi level but show no $k_z$ dispersive pattern. By comparison of the Fermi momenta of the observed TSS in Fig. \ref{Bands}, we ascertain that they are the metallic TSS of the system. Figs. \ref{kz_TSS}(e) and (f) depict the DFT bulk bands at the $(11\bar{2}0)$ surface, in which no vertical band is present. This observation proves that the (0001) projection of the $(11\bar{2}0)$ bands consists of both the bulk electronic states and the TSSs.

\begin{figure}[b]
\centering
\includegraphics[width=9cm]{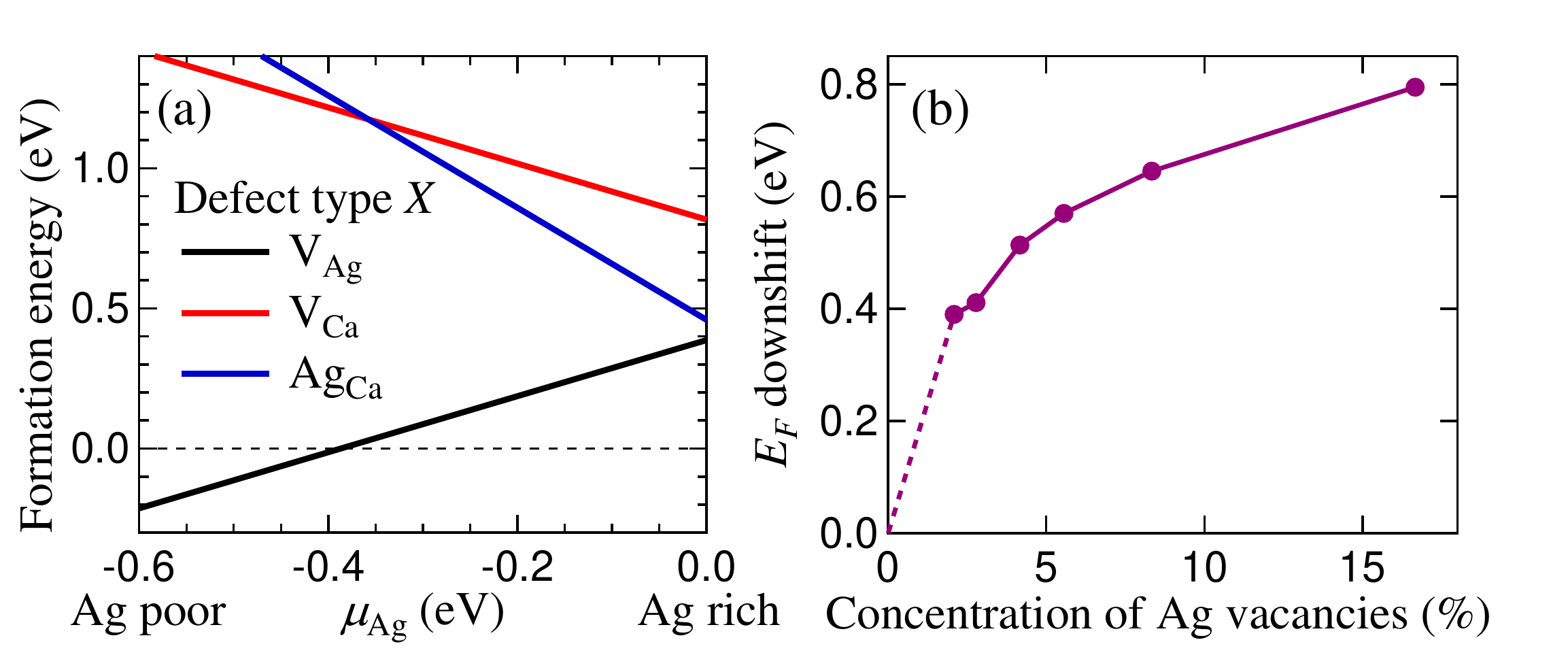}
\caption{(Color online) (a) Formation energies of cation defects. It is clear that Ag vacancies ($\mathrm{V}_{\mathrm{Ag}}$) have the lowest formation energy and are thus energetically favored in natural CaAgAs crystals. (b) Estimation of Fermi level downshift as a function of Ag-vacancy concentration.}\label{Deformation}
\end{figure}

An important discrepancy between the ARPES and the theoretical electronic structure is the position of the Fermi level (carrier density). It is clear from Figs. \ref{Bands}-\ref{kz_TSS} that our as-grown crystals are significantly $p$-doped compared to the perfect lattice used in the DFT \mbox{calculations}. The experimental chemical potential is about 0.5 eV lower than the theoretical one, as demonstrated by yellow dashed lines in Figs. \ref{Bands}(d)-(f). Such $p$-type behavior prevents us from observing the actual Dirac cone (which transfers from the drum-head TSS) by ARPES conveniently, and leads to the need of gating when utilizing this system at its topological insulating regime. In order to understand this behavior, we calculate the deformation energies for different types of lattice defects and estimated the defect concentration needed for certain degree of chemical potential downshift.\cite{Supplement} The calculation result is depicted in Fig. \ref{Deformation}. It is clear from Fig. \ref{Deformation}(a) that Ag vacancies (black line) have the lowest formation energy in both the Ag poor and Ag rich circumstances. Ag vacancies are thus energetically favored in natural CaAgAs crystals. We further see from Fig. \ref{Deformation}(b) that Ag vacancies are very efficient $p$-dopants in the system. It is estimated\cite{Supplement} that a mere 2\% of Ag vacancies (i.e., CaAg$_{0.98}$As) would suffice for a chemical potential downshift of $\sim$0.4 eV. This is consistent with the result of detailed structural analysis on CaAgAs\cite{Private} which sets an upper limit of $\sim$2\% Ag vacancies in the as-grown crystals. In order to obtain $n$-typed crystals, selective bulk/surface electron dopants are needed in the sample growth process, which warrants future investigations.

In conclusion, we performed systematic high resolution ARPES measurements and DFT calculations on candidate nodal-line semimetal CaAgAs. Our ARPES measurements observe a metallic TSS coexisting with a circular bulk band at the Fermi level enclosing $\Gamma$, proving the topological nontrivial nature of this system. Our DFT calculation reveals that CaAgAs is an ideal NLSM with a closed nodal ring near the theoretical $E_F$ when SOC is absent, whereas being a small gap topological insulator with typical Dirac cone TSS when SOC is taken into account. Comparing with the DFT results, we found by ARPES experiments that the as-grown crystals are largely $p$-doped, with a rigid-band-like downshift of $E_F$ of about 0.5 eV. According to our defect-formation energy calculations, such $p$-doping due most likely to dilute Ag vacancies that are energetically favored in the single crystals.

We thank Ling-Yuan Kong, Bo-Qing Lv, Yao-Bo Huang, Ying Zou and Hong Ding for useful \mbox{discussions} and technical supports at the ``Dreamline'' beamline at SSRF. C. L. was supported by the \mbox{National} Natural Science Foundation of China (NSFC) (No. 11504159), NSFC Guangdong (No. 2016A030313650), the Guangdong Innovative and Entrepreneurial Research Team Program (No. 2016ZT06D348), and the Technology and Innovation Commission of Shenzhen \mbox{Municipality} (No. \mbox{JCYJ20150630145302240}). N. N. was \mbox{supported} by the U.S. Department of Energy (DOE), Office of Science, Office of Basic Energy Sciences under Award Number DE-SC0011978. L. H. was supported by the \mbox{NSFC} under Grant No. 11404160, the Shenzhen Key Laboratory Grant No. ZDSYS20141118160434515, and the Shenzhen Peacock Plan Team under Grant No. \mbox{KQTD2016022619565991}. G. W. was supported by The State Key Laboratory of Low-Dimensional Quantum Physics (No. KF201602) and the Technology and Innovation Commission of Shenzhen Municipality (No. ZDSYS20170303165926217, JCYJ20160531190254691).

X.-B.~W. and X.-M.~M. contribute equally to this work.

\end{document}